# Measurement and characterization of infrasound from a tornado producing storm[a]

Brian R. Elbing[1,b], Christopher E. Petrin[1], and Matthew S. Van Den Broeke[2]

[1]Mechanical & Aerospace Engineering, Oklahoma State University, Stillwater, 74078, USA
[2]Earth and Atmospheric Sciences, University of Nebraska-Lincoln, 68588, USA

## ABSTRACT

A hail-producing supercell on 11 May 2017 produced a small (EFU) tornado near Perkins, Oklahoma (35.97, -97.04) at 2013 UTC. Two infrasound microphones with a 59-m separation and a regional Doppler radar station were located 18.7 km and 70 km from the tornado, respectively. Elevated infrasound levels were observed starting 7 minutes before the verified tornado. Infrasound data below ~5 Hz was contaminated with wind noise, but in the 5-50 Hz band the infrasound was independent of wind speed with a bearing angle that was consistent with the movement of the storm core that produced the tornado. During the tornado, a 75 dB peak formed at ~8.3 Hz, which was 18 dB above pre-tornado levels. This fundamental frequency had overtones (18, 29, 36, and 44 Hz) that were linearly related to mode number. Analysis of a larger period of time associated with two infrasound bursts (the tornado occurred during the first event) shows that the spectral peaks from the tornado were present from 4 minutes before to 40 minutes after tornadogenesis. This suggests that the same geophysical process(es) was active during this entire window.

## I.  INTRODUCTION

Numerous natural and anthropogenic sources emit infrasound, sound at frequencies below human hearing (< 20 Hz). Known sources include severe storms (Jones & Georges, 1976; Talmadge & Waxler, 2016), earthquakes (Young & Greene, 1982; Le Pichon et al., 2005; Mutschlecner & Whitaker, 2005), explosions/rocket launches (Waxler et al., 2015; Blom et al., 2016), ocean waves (Waxler & Gilbert, 2006), and volcanoes (Johnson & Ripepe, 2011). Due to weak atmospheric absorption at low frequency and an "acoustic ceiling" within the atmosphere (Bedard & Georges, 2000), infrasound can be detected over significantly larger distances than audible sound. The infrasound carries information about the source, its location/movement, and the environment it passes through between the source and receiver. While this makes infrasound an appealing source for long-range, passive detection and monitoring of infrasound producing events (including tornadoes) as well as the environment it passes through, it makes identification and isolation of a specific source difficult. This is particularly true for tornadoes since they are rare, singular events, and their locations are unknown until minutes before formation. The current work reports field results from infrasound measurements located ~19 km from a verified tornado. A strong infrasound signal was received during the tornado, and the focus of this paper is assessing the likelihood that the received signal was associated with the tornado.

Georges (1973) notes that "the history of the discovery of severe-weather infrasound is clouded by an almost complete absence of early published results." Besides a few publications (Goerke & Woodward, 1966; Bowman & Bedard, 1971), documentation was primarily from internal reports or records of oral conference presentations. Unfortunately, this trend has not changed since these early findings. There were a few related publications in the 1970s (McDonald, 1974; Georges & Greene, 1975; Arnold et al., 1976) before the early 2000s when more activity focused on tornado infrasound. Contemporary work includes several oral presentations (Rinehart, 2012; Goudeau et al., 2018; Elbing et al., 2018), conference papers (Noble & Tenney, 2003; Prassner & Noble, 2004; Bedard et al., 2004a,b), and a recent project report (Rinehart, 2018); but only a few journal articles (Bedard, 2005; Frazier et al., 2014; Dunn et al., 2016).

     Bedard (2005) used an infrasonic observatory and collocated radar to track a velocity couplet aloft that evolved into a tornado and showed maximum circulation descending for ~30 min. The detected infrasound at ~1 Hz followed the trend of the radar observations. In addition, Bedard (2005) notes that a reexamination of an archive of atmospheric infrasound recordings resulted in the identification of over 100 cases with infrasonic signals produced at the time and in the direction of vortices, though few details of these 100 cases have been published. Frazier et al. (2014) examined high-fidelity acoustic recordings covering the frequency range from 0.2 to 500 Hz from three tornadoes in Oklahoma. Primary findings from this work are the use of beamforming at infrasound frequencies to track long-duration tornadoes, detection of audible frequency sound, and demonstration of a modified aeroacoustic jet turbulence model to predict the observed signature in the audible frequency range. This work is further discussed further subsequently, particularly in reference to potential infrasound production mechanisms. Dunn et al. (2016) used a ring laser interferometer to detect infrasound from an EF4 tornado in Central Arkansas on 27 April 2014. Associated infrasound was observed 30 min before the tornado was initially reported and had a fundamental frequency of 0.94 Hz. This is consistent with the observation of Bedard (2005) that large tornadoes produce infrasound in the 0.2-1 Hz range. Three additional vortices that ultimately produced tornadoes were claimed to have been detected at least 30 min before reported touchdown. Thus there is strong evidence that infrasound is produced by a tornado (including during formation), but relatively few observations are well documented in the literature.

Given the dearth of detailed observations of tornado infrasound in the archival literature, the aim of the current work is not to attribute the infrasound observations to a specific tornado mechanism. A coherent understanding of the general mechanism(s) associated with infrasound production from tornadoes will require a broader sampling of infrasound from tornadoes. Consequently, the current objective is to establish confidence that the received signal was associated with the reported tornado and provide sufficient characterization of the storm and received infrasound such that it can be used to test proposed mechanisms. This includes a discussion of proposed mechanisms and whether they are consistent with available observations. The remainder of the paper includes (§2) characterization of the storm and tornado, (§3) analysis of infrasound during the tornado, (§4) discussion and analysis, and (§5) conclusions.

## II.   STORM AND TORNADO CHARACTERIZATION

### *A.   Overview*

On 11 May 2017 a line of storms to the west of the infrasonic array included a hail-producing supercell. At 2013 UTC the supercell produced an EFU tornado (the "U" indicates an unknown rating) near Perkins, OK (35.97, -97.04), which was located 18.7 km south-by-east (SbE) of infrasound microphones at Oklahoma State University (OSU). The official tornado path length and damage width were 0.16 km (0.10 miles) and 46 m (150 ft) (NOAA, 2017), respectively. There were live news reports of a possible second tornado after the first, but it was never confirmed due to the storm being rain wrapped with no low-level radar coverage. Confirmed hail events during the life of the supercell that produced the tornado are provided in Table I, which includes the hail size, UTC time, time relative to the reported tornado touchdown ($t_r$), latitude (lat), longitude (lon), distance between event and the array ($L$), and the bearing angle ($\varphi$) measured from the source to the receiver clockwise relative to north (0°). The largest confirmed hail was 108 mm (4.25 in) that was reported 25 km to the southwest of the infrasonic array at 1956 UTC. There were two reports of hail approximately an hour after the tornado with both events being east of the infrasound microphones.

Table I. Confirmed hail and tornado events within 100 km of the infrasound array during the life of the storm that produced the tornado on 11 May 2017. For hail, the size is the reported diameter in millimeters.

| Event | Size | Time (UTC) | $t_r$ (min) | Lat | Lon | L (km) | $\varphi$ (°) |
|---|---|---|---|---|---|---|---|
| hail | 25 | 19:25 | -48 | 35.95 | -97.59 | 50.2 | 65.7 |
| hail | 38 | 19:38 | -35 | 35.84 | -97.41 | 44.2 | 42.0 |
| hail | 38 | 19:40 | -33 | 35.88 | -97.39 | 39.7 | 44.4 |
| hail | 51 | 19:45 | -28 | 35.95 | -97.28 | 27.2 | 41.0 |
| hail | 44 | 19:50 | -23 | 35.95 | -97.25 | 25.5 | 36.4 |
| hail | 70 | 19:51 | -22 | 35.95 | -97.26 | 26.1 | 38.0 |
| hail | 70 | 19:53 | -20 | 35.95 | -97.25 | 25.5 | 36.4 |
| hail | 44 | 19:56 | -17 | 35.95 | -97.26 | 26.1 | 38.0 |
| hail | 108 | 19:56 | -17 | 35.95 | -97.24 | 25.0 | 34.7 |
| hail | 64 | 20:06 | -7 | 35.84 | -97.25 | 36.1 | 24.8 |
| tornado | EFU | 20:13 | 0 | 35.97 | -97.04 | 18.7 | −11.5 |
| hail | 22 | 21:13 | 60 | 36.22 | -96.57 | 46.7 | −101.8 |
| hail | 19 | 21:15 | 62 | 36.12 | -96.58 | 45.1 | −88.3 |

## B. Ground-level atmospheric conditions

Ground-level atmospheric conditions were monitored by Oklahoma Mesonet stations (Brock et al., 1995; McPherson et al., 2007) and a weather station (termed DML) located ~170 m south of the infrasonic array. The DML weather station (Vantage Pro, Davis Instruments) was located on a building roof and provided 30 minute averages of temperature, humidity, atmospheric pressure, and wind speed. The Oklahoma Mesonet network consists of 120 automated environmental monitoring stations that measure air temperature 1.5 m above ground, relative humidity 1.5 m above ground, wind speed and direction 10 m above ground, barometric pressure, rainfall, incoming solar radiation, and soil temperature. Data are packaged in 5 min "observations" that are quality checked by the Oklahoma Climatological Survey prior to being released. The current study used three sites; Perkins (PERK), Stillwater (STIL), and Marena (MARE).

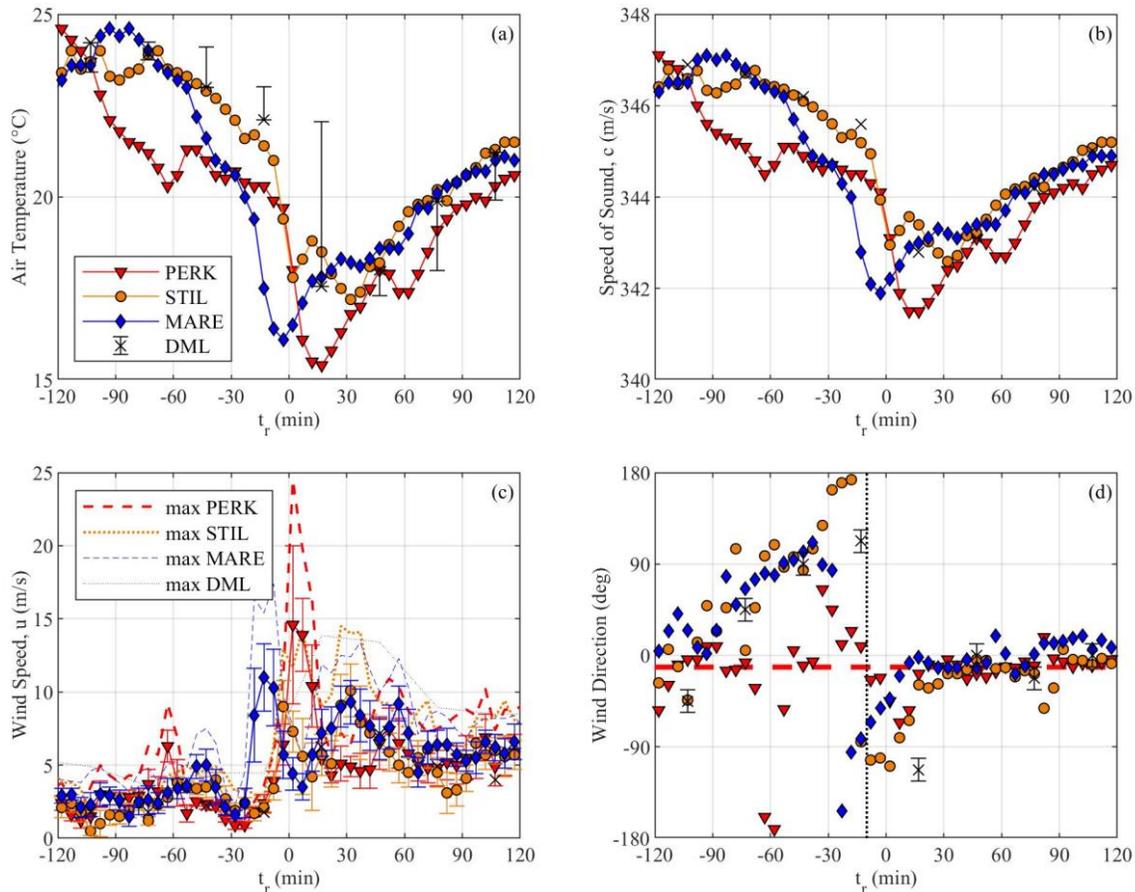

Figure 1. Ground level atmospheric state from the Mesonet and DML weather stations including: (a) Average air temperature with DML error bars being the maximum deviations from the average, (b) speed of sound in humid air, (c) wind speed including maximum 3-second observations, and (d) the wind direction. (Color online)

The Perkins (PERK), Stillwater (STIL), and Marena (MARE) Mesonet stations were located 3.3 km north-by-west (NbW) of the tornado (15.4 km to array), 17.5 km NbW of the tornado (1.9 km to array), and 18.8 km northwest-by-west of the tornado (14.1 km to array), respectively. These stations, in addition to the DML weather station, were used to characterize the ground-level atmospheric conditions and corresponding speed of sound. The measured air temperature, computed speed of sound in humid air (Cramer, 1993), wind speed, and wind direction from each site for the two hours before and after the tornado are provided in Figure 1 with the time ($t_r$) shown relative to the tornado report (11 May 2017, 2013 UTC). The general trends between the sites are all comparable with the primary differences of note being that the minimum temperature occurred earlier at MARE and higher wind speeds were observed at PERK. The higher wind speed was due to the close proximity of the PERK site to the tornado producing storm, and the earlier temperature drop at MARE was due to it being farther west (i.e. the storm reached this site earlier). Of particular note is that the wind direction at PERK from ~10 minutes before the tornado (marked with a vertical dashed line in Figure 1d) through the life of this storm was aligned with the direction from the tornado to the infrasound array (horizontal dashed line in Figure 1d).

## *C.*  *Radar analysis*

Data were analyzed from the Weather Surveillance Radar-1988 Doppler (WSR-88D) at Oklahoma City, Oklahoma (KTLX; 35.33306, -97.27778), which is located ~70 km southwest of the verified tornado (Figure 2). At this range, the base-scan radar beam height was ~0.95 km above radar level (ARL). Data were analyzed from 1928 UTC on 11 May 2017 (~45 min prior to tornadogenesis) to 2103 UTC, when the storm became too far from the radar for good data quality. Around this time, other nearby storms began to merge with the storm of interest.

Several radar metrics were analyzed through the analysis period for the storm of interest (11 May 2017 from 1928-2045 UTC). Base-scan data (radar reflectivity factor and radial velocity) at an altitude of ~0.88 km ARL near the storm core at 1953 UTC are shown in Figure 3. The velocity difference near the surface and associated with the tornadic vortex cannot be robustly analyzed for this event because of its distance from the radar site. Compounding this problem is the small size of the tornado (estimated width ~50 m at the ground), meaning that radar observations at this distance are incapable of measuring maximum wind speeds toward and away from the radar within the tornado. Instead, maximum radial velocity difference (MRVD) was derived for the low-level mesocyclone at base scan, at an altitude of ~1 km ARL. This storm had a well-defined mesocyclone (Figure 3b), so maximum velocity difference was computed as the difference between the associated maximum inbound and outbound velocities. This analysis was done from 1928-2045 UTC, since beyond this time the storm-radar distance increased to too large of a value for velocity difference values to remain comparable.

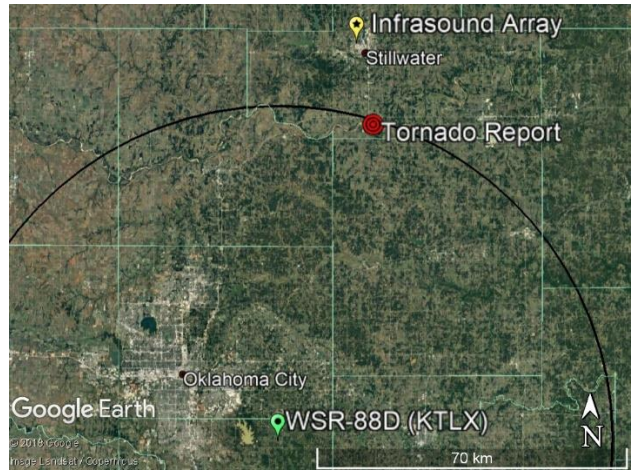

Figure 2. Map showing the location of the WSR-88D radar at Oklahoma City (KTLX). Black circle (radius 75 km) indicates the region where the base-scan beam altitude was less than 1 km ARL, assuming standard beam propagation. (Color online)

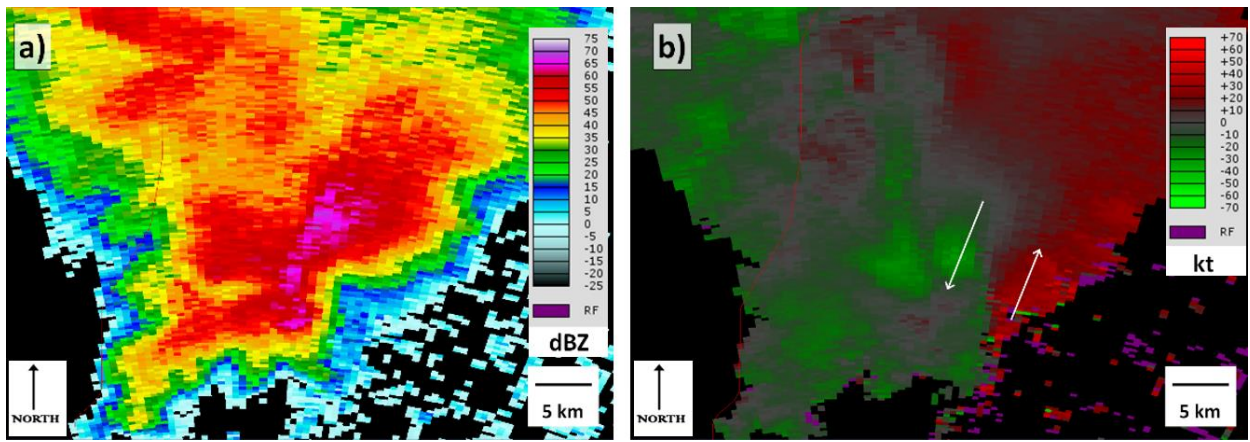

Figure 3. (a) Radar reflectivity factor and (b) radial velocity from KTLX at 1953 UTC on 11 May 2017. Data are base-scan, with an altitude of ~0.88 km ARL near the storm core. Arrows in (b) indicate the inbound and outbound velocities associated with the low-level mesocyclone. (Color online)

Through the analysis period, the MRVD in the low-level mesocyclone ranged from 13.5 m/s to 42.5 m/s (Figure 4). The mesocyclone was relatively weak for a few time steps after initiating, but was well-defined and reasonably strong by 1940 UTC (Figure 4 – MRVD). The low-level mesocyclone reached its maximum intensity at 2000 UTC, ~13 min prior to reported tornadogenesis. Shortly after tornadogenesis, the intensity of the low-level mesocyclone decreased sharply and did not recover during the analysis period. Radar beam-centerline altitude did not change substantially with the low-level mesocyclone intensity, indicating that the observed MRVD changes were genuine changes to storm organization and not an effect of radar beam propagation.

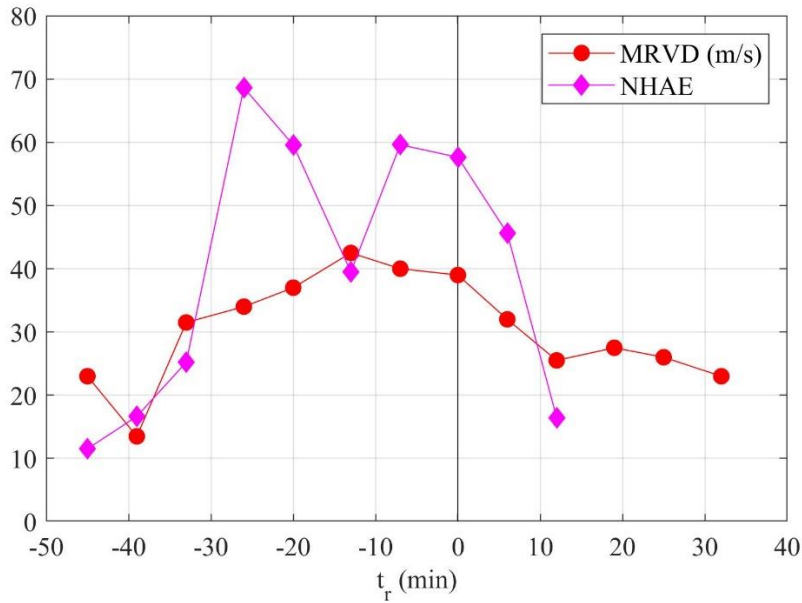

Figure 4. Time series of maximum radial velocity difference (MRVD) and of normalized hail areal extent (NHAE) from 1928-2025 UTC. NHAE has been multiplied by 500 to fit on the same scale. (Color online)

The base-scan normalized hail areal extent (NHAE) > 35 dBZ in radar reflectivity factor ($Z_{HH}$) (Van Den Broeke, 2017) was also analyzed. NHAE uses a combination of $Z_{HH}$ and differential reflectivity ($Z_{DR}$; e.g., Doviak & Zrnić, 2006) to identify areas where hail is present. NHAE is normalized by the storm area, which makes it a percentage of the base-scan storm area

> 35 dBZ dominated by hail and allows for comparison between storms. The time history of the normalized hail areal extent (NHAE) (multiplied by 500 so the magnitude is comparable to MRVD) is also shown in Figure 4. An initial burst of hail around 1947 UTC ($t_r \approx -26$ min) is followed by a secondary burst of hailfall from 2006-2013 UTC ($-7$ min $< t_r < 0$ min). Prior work has noted that hailfall is often maximized in the minutes leading up to tornadogenesis (Browning, 1965; Van Den Broeke et al., 2008). Of note, the majority of the reported hail for this storm occurred in the time spanning these two hailfall bursts with the largest reported hail occurring at $t_r$ = -17 min (Table I). Area of the storm dominated by hail decreased markedly after tornado demise (Figure 4).

## III.   INFRASOUND DURING THE TORNADO

### *A.*   *Infrasound data acquisition*

A 3-microphone (Model 24, Chaparral Physics) infrasonic array was deployed on the campus of OSU during the 2017 tornado season. This effort was part of the CLOUD-MAP project (Elbing & Gaeta, 2016; Smith et al., 2017; Jacob et al., 2018), a multi-university collaboration focused on the development and implementation of unmanned aerial systems (UAS) and their integration with sensors for atmospheric measurement. The infrasonic array, satellite image shown in Figure 5, was centered at (36.1344, −97.0815) and the coordinates for each microphone as well as the separation distances are provided in Table II. Tornadoes generally produce infrasound between 0.5 Hz ($\lambda \approx 686$ m) and 10 Hz ($\lambda \approx 34.3$ m), where $\lambda$ is the acoustic wavelength. Bedard (1998) recommended a nominal spacing of $\lambda/4$ between microphones in an array, though $\lambda/2$ is more widely accepted. Using the half-$\lambda$ spacing, the ideal spacing between microphones is 343 m. Space limitations resulted in the final spacing of ~60 m, which makes it tuned to ~3 Hz (half-$\lambda$ spacing). Each microphone had a nominal sensitivity of ~400 mV/Pa and a nearly flat response from 0.1-200 Hz. All the microphones had identical mounting structure that included a low-frequency vibration isolation pad with the microphone sealed within an acrylic dome painted white. Windscreens were produced using four 15-m long porous hoses connected to each microphone for spatially averaging to cancel out incoherent noise (e.g., wind). The microphones with and without the windscreens (hoses) were tested in an anechoic chamber (though not anechoic to infrasound frequencies) in a method similar to that of Hart & McDonald (2009). These results showed significant reduction in wind noise below 50 Hz without significant attenuation of a reference signal, but no noise reduction by 100 Hz (Threatt, 2016). The microphones were powered with DC-power supplies (APS-1303, Aktakom). The output from each microphone was recorded via a dynamic signal analyzer (USB-4432, National Instruments). The data acquisition was controlled via a commercial software package (Sound & Vibration Measurement Suite, National Instruments). The sample rate was fixed at 1 kHz and grouped in 20-minute observation windows. Unfortunately, there was cross-talk between microphones 2 and 3 that was not identified until after the reported observation. However, some analysis between microphones 1 and 3 is provided.

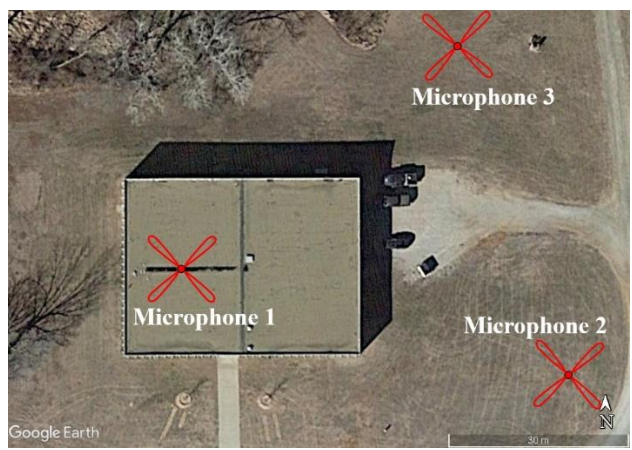

Figure 5. Satellite image of the OSU infrasonic array. Microphone locations are labelled and denoted by the 'X'. (Color online)

Table II. Summary of locations for each of the microphones, mounting location (roof or ground level), and the separation distance between the microphones.

|       | **Location** | | | **Separation Distances (m)** | | |
|---|---|---|---|---|---|---|
|       | **Latitude** | **Longitude** | **Mounting** | **Mic 1** | **Mic 2** | **Mic 3** |
| **Mic 1** | 36.1344 | −97.0819 | roof | 0 | 67.6 | 58.6 |
| **Mic 2** | 36.1342 | −97.0812 | ground | 67.6 | 0 | 58.5 |
| **Mic 3** | 36.1347 | −97.0814 | ground | 58.6 | 58.5 | 0 |

## *B.    Time series analysis*

For the current analysis microphone 3 data is included in spite of the cross-talk problem since it contains independent data (though contaminated with "noise" from microphone 2). The time trace of microphones 1 and 3 are provided in Figure 6 with the time ($t_r$) relative to the tornado report (11 May 2017, 2013 UTC). It is apparent that a signal that was significantly stronger than the background levels peaks close to the time of the tornado report, and a second strong event begins ~20 minutes after the reported tornado touchdown. While seemingly aligned with the tornado report (and a potential, un-confirmed rain wrapped second tornado), inspection of the wind speed (STIL Mesonet) time trace (included in Figure 6) shows that the elevated infrasound levels are correlated with when the wind speed exceeds ~5 m/s.. This is consistent with Pepyne & Klaiber (2012) that observed that porous hose filters were not effective windscreens when the wind exceeds 5 m/s. This is also consistent with the fact that microphone 3 (as well as microphone 2, not shown) was attenuated relative to microphone 1 since microphone 1 was elevated and microphones 2 and 3 are on the ground with some natural wind breaks (e.g. trees, buildings) surrounding them. In addition, there is a possibility that the porous hoses on microphones 2 and 3 were infiltrated by rain water due to being laid directly on the ground. Conversely, microphone 1 was on a roof with good drainage and the porous hoses were elevated 38 mm above the surface. For these and other issues (Hart & McDonald, 2009; Pepyne & Klaiber, 2012) there is active research searching for alternative windscreen options (e.g. close-cell dense foam; Zuckerwar, 2010;

Shams et al., 2005; 2013; Alberts et al., 2013; Dauchez et al., 2016) with most infrasound researchers moving to the use of porous domes (Talmadge, 2018).

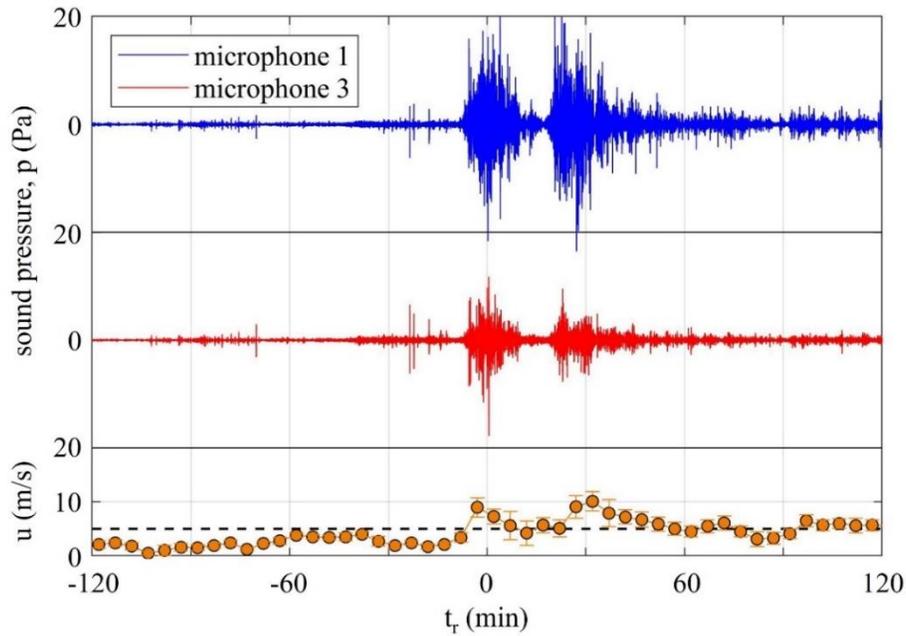

Figure 6. The measured sound pressure versus time in minutes relative to the tornado touchdown (11 May 2017, 2013 UTC) for microphones 1 and 3 with the amplitude of microphone 3 shifted for clarity. The time trace of the STIL Mesonet site with a reference dashed line at 5 m/s is also included for comparison. (Color online)

These infrasound signals are nonstationary, but for the purpose of analysis such signals can often be viewed as piecewise stationary. This requires a means of identifying the appropriate period, which is challenging since this is a singular event (i.e. a unique tornado with additional, unknown background infrasound sources). Here the analysis method of Bendat & Piersol (2000) for a nonstationary single record is followed with the assumption that the single measurement is the product of a deterministic function and a random process. This analysis demonstrates that increasing the averaging period $T$ reduces the random errors but increases the bias error. Thus the selection of the period is critical for an accurate representation of the data during the analysis. For the current work, the appropriate averaging period was determined from the trial-and-error approach (Bendat & Piersol, 2000). Figure 7 shows the squared effective pressure, $P_e^2 = \frac{1}{T}\int_0^T p^2 dt$, from microphone 1 with averaging periods from 0.01 s to 1000 s (additional periods were examined, though not shown). From these results it is clear that $T = 0.01$ s and 10 s still have abrupt variations from one sample to the next, which is indicative of random errors. Conversely, the results of $T = 1000$ s shows a significant bias error as illustrated from the observation that $P_e^2$ increases before the actual infrasound signal rises as $t_r = -7$ min and 20 min. Thus from trial-and-error a final averaging period of $T = 100$ s was determined. Consequently, subsequent data analysis was performed within 100 s windows.

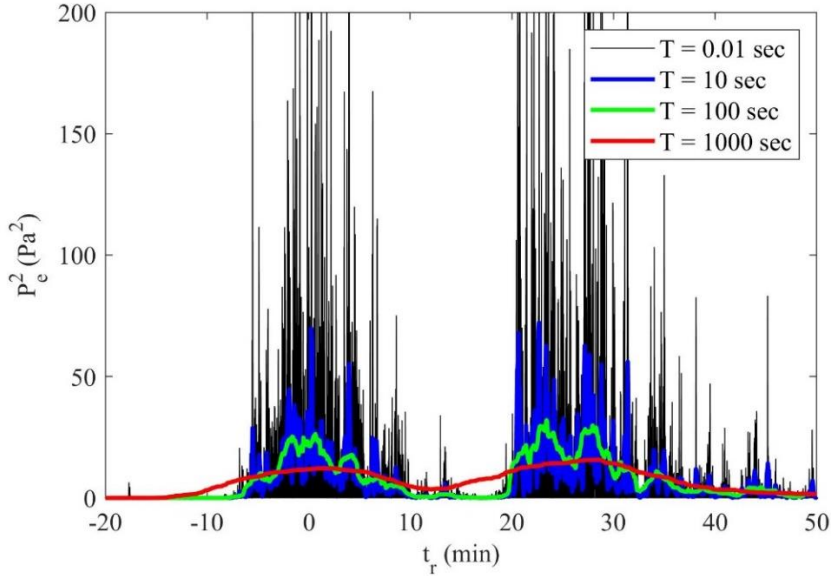

Figure 7. Squared effective pressure ($P_e^2$) versus time with a wide range of averaging periods. The longest period ($T = 1000$ sec) creates a large bias error as evident at $t_r \sim 17$ min. (Color online)

## C. *Spectral analysis*

The sound pressure spectra, $\Phi(f)$, presented herein are the single-sided form such that

$$P_{rms}^2 = \int_0^\infty \Phi(f) df, \tag{1}$$

where $P_{rms}^2$ is the pressure variance and *f* is the temporal frequency. For the current analysis, the period of time when the tornado was present was set at $-46 < t_r < 154$ s, which corresponds to

±100 s from the nominal time arrival for the direct path acoustic wave ($t_r = 54$ s). The ±100 s window was selected because it is consistent with the averaging period previously determined as well as being nominally consistent with the level accuracy with which it is known the tornado was present. To determine the pressure spectra during the tornado, the period of interest ($-46 < t_r < 154$ s) was segmented into 100 sec periods with 75% overlap. The square of the double-sided fast Fourier transform (FFT) was multiplied by two to give the single-sided sound pressure spectra (i.e. power spectral density). The accuracy of each spectrum was checked against equation (1), and the variation was less than 1% for all computed spectra. All of the individual spectra within the window of interest were averaged to provide the mean spectrum. Results from microphone 1 (individual segments as well as the mean spectrum) during the tornado are provided in Figure 8 with the sound pressure spectra reported in decibels referenced to 20 µPa. Here there is gradual decay in the power spectral density from ~0.1 Hz until a broad peak is observed between 5 and 14 Hz. The elevated spectral levels below 5 Hz are likely due to wind noise, which is supported by the infrasound amplitude at these low frequencies being well correlated with the local wind speed. The broad peak was smoothed with a 1/50$^{th}$ decade filter and then the maximum energy was used to identify the peak of 75 dB at 8.3 Hz with a quality factor of 1.6. The quality factor $\left(Q = f_{pk}/(f_U - f_L)\right)$ is a measure of the bandwidth of the peak, where $f_U$ and $f_L$ are the frequency 3 dB below the peak on the upper and lower side of the peak, respectively. Following the initial peak at ~8.3 Hz, there are overtones with nominal peaks at 18, 29, 36, and 44 Hz, which have quality factors ($Q$) of 3.4, 3.8, 4.8, and 4.9, respectively. After the last overtone there is a rapid roll-off associated with the low-pass filter created by the porous hose windscreens. The fundamental frequency (8.3 Hz) peak was ~18 dB above the levels before the rise in infrasound associated with the tornado ($-27 < t_r < -7$ min).

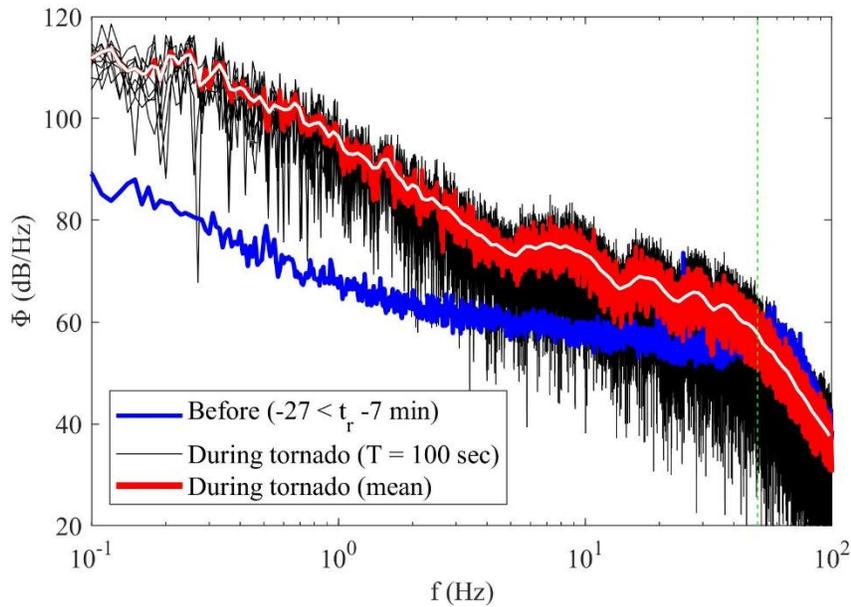

Figure 8. Sound pressure spectra during the nominal time of arrival of signals emitted from the verified tornado (−46 < $t_r$ < 154 s) compared with the sound pressure spectra before the rise in infrasound (blue line). Thin black lines correspond to spectra from individual 100 sec intervals, and the thin white line is the mean with a 1/50$^{th}$ decade filter applied. (Color online)

### *D.* *Bearing angle estimate*

As the larger amplitude content was likely associated with wind noise, the question is whether these other peaks are associated with the tornado. The cross-talk contamination between microphones 2 and 3 prevents bearing angle calculation without applying assumptions about the resulting pressure wave and its orientation. However, it is possible to answer the question of whether the received signals are consistent with what is expected if they were produced by the tornado. For this analysis it is assumed that the distance between the tornado and the array (18.7 km) was sufficient that the received signals are well approximated as plane waves. Given the speed of sound (343.8 m/s; mean from STIL, DML, PERK) and the frequency range of interest (5 < $f$ < 14 Hz), the distance between the array and tornado corresponds to 270 to 760 wavelengths. Next it is assumed that the received signals were propagating parallel to the ground directly from the source. Note that given a range of 18.7 km with possible cross-wind propagation, this assumption adds uncertainty to the calculations.

The bearing angle of the filtered signal was determined with time-domain beamforming using the time difference of arrival technique (Dowling & Sabra, 2015). The separation (or lag) time between microphones 1 and 3 ($t_{13} = t_1 - t_3$) was determined from the peak in the cross-correlation between the two signals. Given the assumptions, there are two valid bearing angles mirrored about the line between microphones 1 and 3 (if the horizontal plane wave assumption were not applied this would represent a cone). The speed of sound (*c*) and the distance between microphones 1 and 3 ($L_{13}$) can be used to define the angle between the plane wave front relative to the line connecting microphones 1 and 3, $\theta = \cos^{-1}(c|t_{13}|/L_{13})$. Then with geometric relationships, the bearing angle of the received signal ($\varphi$) and its mirrored result ($\varphi'$) can be determined with the angle measured positive clockwise from north (0°).

The sensitivity of the bearing angle to the processing parameters showed the largest measurement uncertainty. Thus the bearing angle was computed by applying a 5$^{th}$ order Butterworth bandpass filter with the minimum and maximum cutoff frequencies incrementally varied between 5.5-11.5 Hz and 40-50 Hz, respectively. These frequency ranges were selected because they nominally span the width of the fundamental peak and the 4$^{th}$ overtone, respectively. In addition, the segment period and overlap percentage between segments was varied, which produced a total of 189 computed bearing angles per time step. In Figure 9, the resulting time history of the mean bearing angle for the received infrasound signals is shown with the error bars equal to the standard deviation determined from variation of the processing parameters. Only $\varphi'$ is shown in Figure 9 for clarity since this analysis is simply to show that the directionality of the received signals are consistent with that expected from the tornado. For comparison, the relative bearing angles for the front, middle, and back side of the storm core that produced the tornado are also provided in Figure 9. Here the 'storm core' location was defined via radar reflectivity as the region of the storm that produced the tornado with > 50 dBZ at a nominal elevation of 1 km. In addition, a horizontal reference line at the bearing angle corresponding to the tornado report ($\varphi_{tornado} = -11.5°$) is included, which intersects the storm core center curve at $t_r = 0$. It should be noted that at ~2035 UTC ($t_r \approx 22$ min) the storm core breaks into two segments with the bearing angles shown corresponding to the front and back of the leading and trailing segments, respectively.

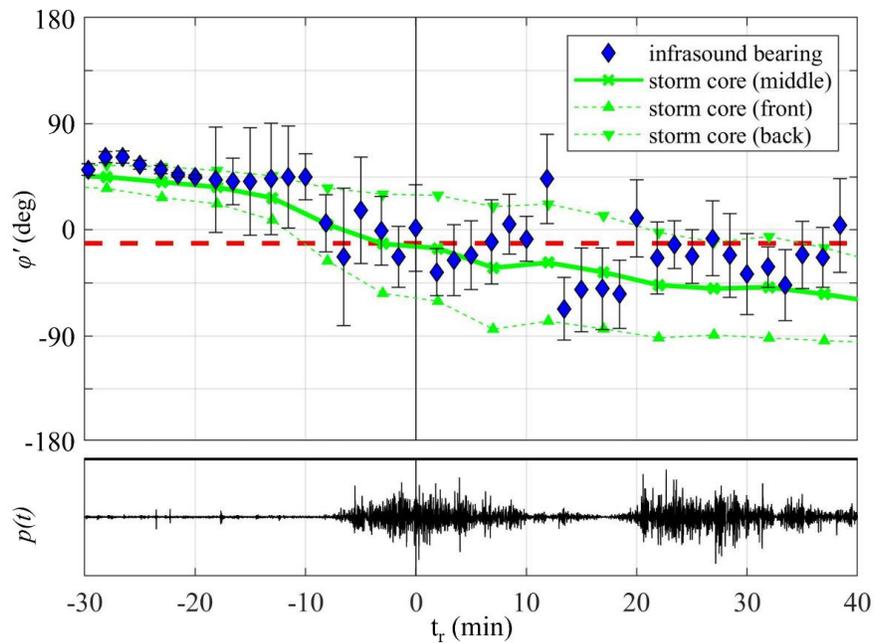

Figure 9. Bearing angle ($\varphi'$) of the received signal bandpass filtered between 5-50 Hz. The horizontal dashed line is the tornado bearing angle. The other dashed lines correspond to the nominal bearing angles of the leading or trailing edges of the storm core while the solid line is the storm core center. Sound pressure time trace is also included for reference. Mirrored bearing angles are omitted for clarity. (Color online)

While there is significant scatter in the results, the mean bearing angles track with the general storm core direction with the majority of data points falling between the bearing angles corresponding to the leading and trailing edges of the storm core. Ultimately, given the applied assumptions and known noise contamination from the other microphone, the bearing angles are nominally consistent with what would be expected if emitted from the reported tornado location. This gives corroborative evidence that the infrasound signal of interest originated from the region within the storm that produced the tornado.

## IV. DISCUSSION AND ANALYSIS

### A. *Characterization of events (infrasound bursts)*

The time trace of the sound pressure (Figure 6) shows that there were two distinct infrasound bursts or events. This is interesting given the possibility of a second tornado that was not confirmed due to the rain wrapped storm and lack of low-level radar. Examination of the sound pressure and the cross-correlation from microphones 1 and 3, $C_{1,3}(t) = \frac{1}{T}\int_0^T P_1(\tau)P_3(\tau + t)d\tau$, showed elevated sound pressure and correlation levels during $-7 \leq t_r \leq 52$ minutes, where $\tau$ is the lag (or shifted) time between signals. A more detailed examination inside of this window shows that the spectral peaks seen in Figure 8 first appear at $t_r = -4$ minutes, and they persist until $t_r = +40$ minutes, including the period between the two bursts where the sound pressure levels were reduced. Figure 10 shows the pressure spectra during event 1 ($-4 < t_r < 11$ min), event 2 ($19 < t_r < 40$ min), the period between events 1 and 2, and the spectra before and after the rise in infrasound.

First, it is important to note that the power spectra during the larger window of event 1 are nearly identical to those of the narrow window used to analyze Figure 8. Furthermore, events 1 and 2 are also nearly identical, which suggests that they are both related to a similar physical process(es). It is also interesting that the relatively quiet period between the two events had pressure spectra that looked similar to events 1 and 2, though at reduced levels. This suggests that the same mechanism was active throughout the period of interest with the source either weakening for a period or the propagation path changing (e.g. wind speed/direction, source elevation, storm structure, etc). Here we note that, while the infrasound community has made great advances in the study of acoustic propagation of infrasound (Ostashev et al., 2005; Le Pichon et al., 2010; Waxler & Assink, 2017), no corrections for propagation effects have been applied for the current work. For $t_r > 40$ min, the tones are lost even though the spectra levels in the 5-50 Hz band remain elevated, but they did drop to comparable levels as before event 1 for frequencies above ~5 Hz. Below 5 Hz the signal remain elevated above the background levels, which is associated with the wind noise and can be seen in Figure 6 with higher wind speeds and sound pressure levels.

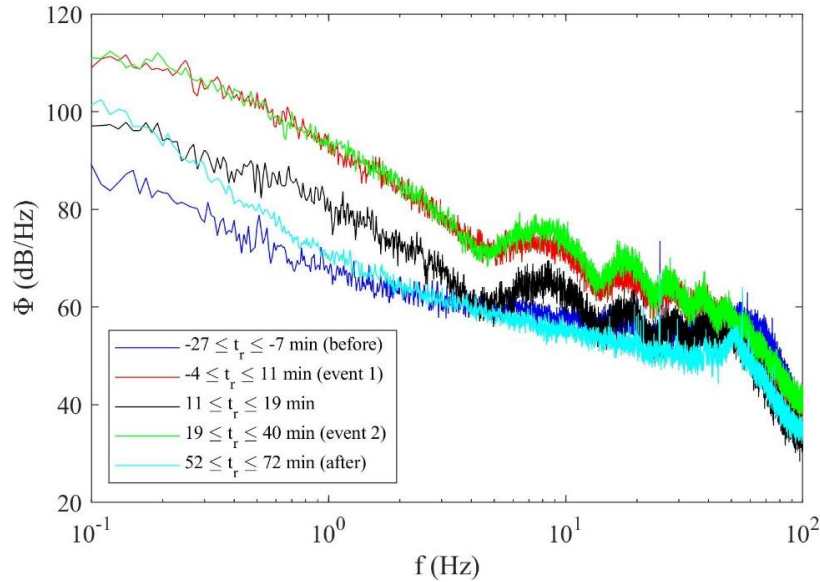

Figure 10. Comparison of the pressure power spectra during the infrasound bursts (events), the period between the events, and for reference the spectra before and after these events. (Color online)

While the fundamental frequency and the associated overtones during the tornado were identified from the power spectra discussion, it is informative to precisely identify the peaks during the various periods of interest. The fundamental and first 4 overtones were identified for periods before, during, and after both infrasound events based on the maximum energy. These results as well as the peak amplitude are provided in Table III. Note that the bands of interest were also integrated and compared with the peak amplitudes, which were nearly identical. Inspection of Table III shows that the frequency of the peaks for the fundamental and overtones were similar between events 1 and 2 with the mean deviation between events being 2%. The amplitudes had a mean deviation of 3%, but consistently event 2 had the higher amplitude. As previously mentioned (and observed in Table III), the period between events 1 and 2 also had peaks at nearly the same frequencies, though with a decrease in amplitude (~6 dB relative to event 1). This gives strong evidence that the mechanisms leading to the two events were related.

Table III. Frequency and amplitude of the fundamental ($n = 0$) and overtones ($n > 0$) from before, during, between events, and after both infrasound events.

| $N$ | 0 | 1 | 2 | 3 | 4 |
|---|---|---|---|---|---|
| Frequency Band, a-b (Hz) | 5-14 | 14-23 | 23-32 | 32-43 | 43-52 |
| **Before Tornado(es)** ($-27 < t_r < -7$ min) | | | | | |
| $f_n$ (Hz) | 13.8 | 15.2 | 25.1 | 41.7 | 50.2 |
| Peak (dB/Hz) | 57.3 | 57.7 | 57.4 | 56.4 | 57.1 |
| **Infrasound Event 1** ($-4 < t_r < 11$ min) | | | | | |
| $f_n$ (Hz) | 8.3 | 18.2 | 27.6 | 38.1 | 45.8 |
| Peak (dB) | 73.1 | 66.2 | 63.5 | 61.3 | 58.7 |
| **Between Events** ($11 < t_r < 19$ min) | | | | | |
| $f_n$ (Hz) | 8.3 | 19.1 | 27.6 | 38.1 | 47.9 |
| Peak (dB) | 66.0 | 59.4 | 56.7 | 56.2 | 56.8 |
| **Infrasound Event 2** ($19 < t_r < 40$ min) | | | | | |
| $f_n$ (Hz) | 8.7 | 18.2 | 27.6 | 38.1 | 47.9 |
| Peak (dB) | 76.0 | 69.7 | 65.1 | 62.0 | 59.2 |
| **After Tornado(es)** ($52 < t_r < 72$ min) | | | | | |
| $f_n$ (Hz) | 5.3 | 19.6 | 25.0 | 39.1 | 51.2 |
| Peak (dB) | 60.2 | 60.5 | 66.9 | 55.4 | 56.1 |

Since Abdullah (1966) predicted overtones,

$$f_n = \frac{(4n + 5)c}{4d}, \qquad (2)$$

it is informative to compare the current observations with these predictions. Here *n* is a non-negative integer, and *d* is the diameter of the vortex core. Abdullah (1966) produced this relationship by modelling a tornado as a compressible Rankine vortex, which when constrained to axisymmetric vibrations with a large vertical-to-radial wavelength ratio. Given the observed fundamental frequency from the current observation, the resulting overtones are compared with the predictions from equation (2). It is important to remind readers at this point that Schecter (2012) identified several fundamental issues with this analysis that precludes this mechanism, which are discussed in more detail subsequently. Thus it is not surprising that in Figure 11 there is a significant deviation between the current observations and that predicted by Abdullah (1966). The error bars on the current results are set based on the quality factor (i.e. 3-dB reduction on each side of the peak). It is interesting that the observed overtones are linearly related ($f_n = 9.47n + 8.64$) but not pure harmonics (factor of ~1.1 rather than 1.0 between overtones). Linear regression analysis shows that both the linear slope (9.47±0.56) and intercept (8.64±1.38) were statistically significant (p-value < 0.05). Furthermore, using a t-test on the slope shows that the multiplication factor between overtones was between 1.07 and 1.21 with 95% confidence. This is a potentially important observation with respect to identifying a fluid mechanism for the infrasound production, but more tornado observations are required to explore these relationships in greater detail.

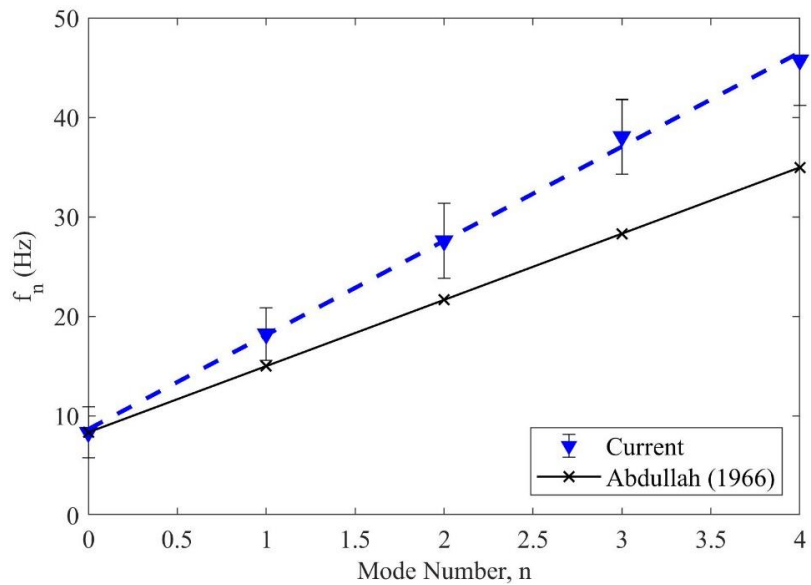

Figure 11. Fundamental and overtones during infrasound event 1 (tornado) compared with the predictions from equation (2) given the fundamental frequency. (Color online)

## *B.  Comparison with potential mechanisms*

While the current work does not aim to attribute the observations to a specific mechanism, it can be informative to compare the current observations to proposed mechanisms. First, it should be noted that the higher frequency signature Frazier et al. (2014) observed and demonstrated to be consistent with aeroacoustic jet turbulence was not observed. This is expected since Frazier et al. (2014) measured larger tornadoes (EF-2, EF-4, EF-5) and noted that the frequency range was 10-100 Hz, which means the current small tornado (EFU) would likely have a signature above 100 Hz (i.e. at a frequency higher than the current acoustic array could observe). Frazier et al. (2014) also notes that there was evidence of a lower frequency (< 2 Hz) signature. The low frequency signal was assumed to be mostly due to wind noise, but there was sufficient coherence for successful beamforming to produce bearings to the tornado producing storms. This lower frequency signature has had several proposed mechanisms including radial oscillations (Abdullah, 1966; Bedard, 2005; Schecter, 2012), electromagnetic sources (Balachandran, 1983; Few, 1985; Pasko, 2009), co-rotating vortices (Powell, 1964; Georges, 1976), vortex-surface-interactions (Tatom et al., 1995), heat-related sources (Nicholls et al., 2004; Akhalkatsi & Gogoberidze, 2009; Schecter & Nicholls, 2010; Markowski & Richardson, 2010; Schecter, 2012), and non-equilibrium effects (Zuckerwar & Ash, 2006; Ash et al., 2011).

As previously mentioned, Schecter (2012) has demonstrated that the Abdullah (1966) analysis has fundamental issues; primarily (i) constraints on the tangential velocity fluctuations at *d*/2 are nonphysical, (ii) requirements on outward propagation of acoustic waves are not met, and (iii) the solution includes modes for nonphysical acoustic sources outside of the vortex. In addition to the failure of Abdullah (1966) to predict the overtones of the current observations (Figure 11), Figure 12 compares the Abdullah (1966) predictions against available observations (Bedard 2005;

Dunn et al., 2016; current). The diameter for the current observation was set at the maximum damage path width (46 m). The maximum damage path is not the vortex diameter but is the best measure of the current observation since low-level radar data were not available. There is a low quality video of this tornado, in which the visible part of the funnel cloud has a maximum thickness that was ~2.9 times the width at the ground. This was used to provide a nominal uncertainty estimate for the current tornado diameter. Also, the error bars shown for the fundamental frequency was determined from the 95% confidence interval from the linear regression analysis of the spectral peaks evaluated at $n = 0$. This shows that the current observation is the only measurement that falls on the predicted fundamental curve of Abdullah (1966), but with the tornado size uncertainty biased towards the higher harmonics. All of the other observations tend to align better with the first or second harmonic.

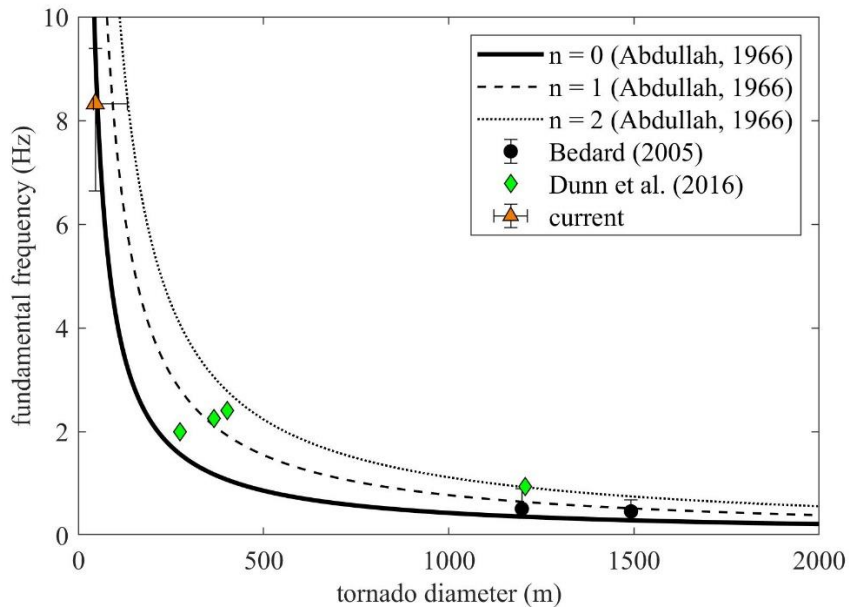

Figure 12. Available observations of infrasound fundamental frequencies associated with tornadoes with estimates of the tornado size compared against predictions from Abdullah (1966).

The prolonged signal that began prior to tornadogenesis is inconsistent with electromagnetic sources (impulsive sources) and vortex-surface-interactions (signals should not be observed prior to touching the ground). In addition, the frequencies of the current and previous observations were too high for those predicted based on co-rotating vortices, and the original postulate that motivated this mechanism has been disproven. Recent simulations (Schecter & Nicholls, 2010; Schecter, 2012) show that there is a lack of discernible infrasound in the absence of latent-heating effects and that non-tornadic thunderstorm cells produce infrasound from the melting level. This suggests that latent heat sources are a likely mechanism, and simulations of the liquid-vapor transitions within a cloud were able to produce infrasound between 0.1 and 10 Hz (Akhalkatsi & Gogoberidze, 2009; Schecter & Nicholls, 2010). However, radial vortex oscillations including the non-columnar nature of a tornado (Schecter, 2012) and analysis incorporating non-equilibrium effects (Zuckerwar & Ash, 2006; Ash et al., 2011) are also consistent with observations. As noted by Frazier et al. (2014), there are likely multiple acoustic generation mechanisms active, which was based on their datasets exhibiting coherent acoustic energy within two distinct regimes (< 2 Hz and 10-100 Hz with larger tornadoes; EF-2, EF-4, and EF-5).

## C.   *Comparison with radar metrics*

Given a history of a complex relationship between hail production, vorticity, and infrasound production (Bowman & Bedard, 1971; Bedard, 2005; Schecter et al., 2008), the current infrasound observations (power spectral peak within the 5-14 Hz band, $\Phi_{max}$) are compared with the radar metric normalized hail areal extent (NHAE) in Figure 13. In addition, the maximum radial velocity difference (MRVD) is included to demonstrate that infrasound production was not correlated with the large-scale rotation of the supercell that produced the tornado (i.e. the MRVD peak occurs before the rise in infrasound). Each parameter in Figure 13 has been normalized to facilitate comparisons. The maximum values used to scale the pressure power spectra ($\Phi_{max}$), MRVD, and the NHAE were 0.3123 Pa$^2$/Hz, 42.5 m/s, and 0.137, respectively. The NHAE during event 1 generally follows the infrasound with their rise, peak, and roll off occurring at nearly the same time. This is consistent with observations by Schecter et al. (2008) that infrasound from a tornado-like vortex radiates infrasound in the 0.1-10 Hz range from the region where diabatic processes involving hail are active. It should be noted, however, that before the infrasound signal was observed significant hail was produced from this storm as evident from both radar metrics and hail reports (Table I). Ultimately these results suggest that infrasound from a tornadic storm could be connected with hail production, but hail production is not solely responsible for the infrasound production.

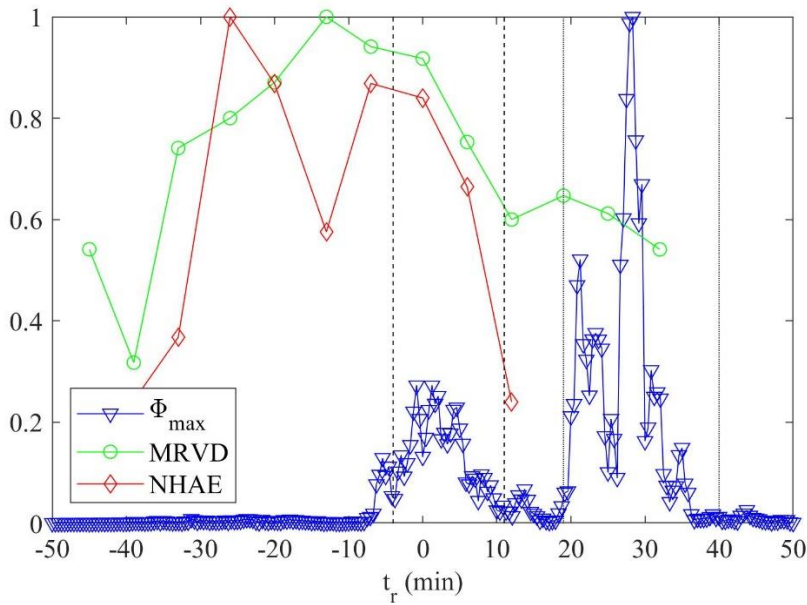

Figure 13. Time histories of the normalized pressure power spectra within the 5-14 Hz band ($\Phi_{max}$), maximum radial velocity difference (MRVD), and the normalized hail areal extent (NHAE). Vertical dashed and solid lines denote the windows corresponding to infrasound event 1 and 2, respectively. (Color online)

## V.    SUMMARY AND CONCLUSIONS

The current work presents infrasound measurements during a hail-producing supercell in Oklahoma on 11 May 2017, which produced an EFU tornado near Perkins, OK (35.97, -97.04) at 2013 UTC with a path length of 0.16 km and damage path width of 46 m. The storm was characterized using ground based measurements (Mesonet sites and a weather station at the infrasound sensors) of the air temperature, humidity, pressure, wind speed, and wind direction. The closest WSR-88D radar (KTLX) was too far (~70 km) to measure maximum wind speed within the tornado, but the MRVD from low-level mesocyclone at ~1 km ARL and NHAE were used to characterize the storm before, during, and after the tornado. There were reports of a possible second tornado after the first, but it was never confirmed due the storm being rain wrapped.

Two infrasound microphones with porous hoses as windscreens were recording 18.7 km from the tornado. The data below ~5 Hz was contaminated with wind noise during the tornado, but the 5-50 Hz band produced data above the noise floor. During the tornado, a fundamental frequency of 8.3 Hz was observed with overtones at 18, 29, 36, and 44 Hz. While two microphones were insufficient to identify a definitive bearing angle, assumptions about the propagation and filtering the data between 5-50 Hz showed that the received infrasound was consistent with that expected from the storm core that produced the tornado. Furthermore, the bearing angle of the received signal during the confirmed tornado was within the uncertainty of the bearing angle measurement.

The spectral peaks observed during the tornado were present from −4 to +40 minutes relative to the confirmed tornado (2013 UTC). The time trace shows two significant bursts, which the events were identified based on cross-correlation between the microphones as event 1 (−8 to +10 minutes) and event 2 (20 to 30 minutes). The power spectra from events 1 and 2 were nearly identical to the narrow period corresponding to the confirmed tornado. In addition, the period between events 1 and 2 showed similar structure in the power spectra, though at a lower amplitude. This suggests that the second event (as well as the period between) had similar active physical processes, and if a second tornado did occur it was likely from the same geophysical process that produced the first tornado. The overtones observed in the spectra were shown to be linearly related, but not matching those predicted by Abdullah (1966). In addition, comparison of the current results (noting that the damage path width likely underestimates the vortex core size) and past observations (Bedard, 2005; Dunn et al., 2016) show Abduallah (1966) fails to predict the tornado size given the fundamental frequency. However, the separation of the available data does support the conjecture that a relationship between tornado size and the infrasound frequency does exist. While no specific mechanism was considered as a potential explanation for the current results, the consistency of the current observation (as well as those in the literature) with various proposed mechanisms were discussed. Electromagnetic sources, vortex-surface-interactions, and co-rotating vortices are inconsistent with observations. While, latent heat effects, radial vortex oscillations that include non-columnar nature of a tornado, and non-equilibrium effects are consistent with observations. In addition, it was noted that the current observations did not measure at sufficiently high frequency (> 100 Hz) to assess the aeroacoustic jet turbulence signature proposed in Frazier et al. (2014).

     Finally, comparison of the infrasound with radar metrics produced insights about the infrasound and its relationship to the larger storm system. The MRVD of the base level of the mesocyclone was not well correlated with the infrasound, which suggests that the large-scale storm rotation is not a mechanism for the infrasound. This is consistent with other observations that the mesocyclone rotation is not responsible for the production of infrasound, but rather the tornado

structure. The normalized hail areal extent (NHAE) did appear to be correlated with the infrasound in the 5-14 Hz band during event 1, which includes the confirmed tornado. Here the rise, peak, and roll off of both the infrasound and NHAE occurred nearly simultaneously. This supports the observation in the literature that the infrasound could be connected with the diabatic processes involving hail activity. However, the maximum observed NHAE for this storm occurred prior to the production of significant infrasound in the 5-14 Hz band, which is consistent with past observations that hail production alone (e.g. without rotation) does not produce infrasound. While it was unfortunate that the radar was too far from this tornado for characterization of the tornado, it does demonstrate the potential use of infrasound to characterize even weak tornadoes in remote locations where low-level radar coverage is poor. This work combined with future observations of tornado infrasound should provide insights into the fluid mechanism(s) responsible for infrasound production. In addition, there is a need for a more detailed analysis of hail production, vorticity, and infrasound production.

## ACKNOWLEDGEMENTS

The authors would like to thank Arnesha Threatt and Shannon Maher, who originally set up and deployed the infrasonic microphones. In addition, we would like to acknowledge the Oklahoma State University infrasound team (Jalen Golphin, Jared Hartzler, Shelby Webb, Alexis Vance, Katrine Hareland, and Logan King) that have worked to maintain the array and inspect the data. This work was supported by the National Science Foundation (NSF) under Grant 1539070: CLOUD-MAP – Collaboration Leading Operational UAS Development for Meteorology and Atmospheric Physics (Timothy VanReken, Program Manager), National Oceanic and Atmospheric Administration (NOAA) under Grant NA18OAR4590307 as well as by B.R. Elbing's Halliburton Faculty Fellowship endowed professorship.